\documentclass[aps,prresearch,twocolumn,superscriptaddress]{revtex4-2}

\usepackage{appendix}
\usepackage{subcaption}
\usepackage{xcolor}
\usepackage{amsmath}
\usepackage{graphicx}
\usepackage{dcolumn}
\usepackage{bm}
\usepackage{hyperref}

\usepackage[normalem]{ulem}   

\begin{document}

\preprint{APS/123-QED}

\title{Operator dynamics in $k$-Markov random circuits}%

\author{Unnati Akhouri}
\affiliation{Institute for Gravitation and the Cosmos, The Pennsylvania State University, University Park, PA 16802, USA}
\affiliation{Department of Physics, The Pennsylvania State University, University Park, PA 16802, USA.}

\author{Pei-Jun Huang}
\affiliation{
Department of Physics, Brown University, Providence, RI 02912, USA}

\author{Elliott Rose}
\affiliation{Institute for Gravitation and the Cosmos, The Pennsylvania State University, University Park, PA 16802, USA}
\affiliation{Department of Physics, The Pennsylvania State University, University Park, PA 16802, USA.}

\author{Sarah Shandera} 
\email[Corresponding author: ]{ses47@psu.edu} 
\affiliation{Institute for Gravitation and the Cosmos, The Pennsylvania State University, University Park, PA 16802, USA}
\affiliation{Center for the Theory of Emergent Quantum Matter, The Pennsylvania State University, University Park, PA 16802, USA}
\affiliation{Department of Physics, The Pennsylvania State University, University Park, PA 16802, USA.}

\date{\today}

\begin{abstract}
We demonstrate that $k$-Markov sequences of unitary gates provide low-cost handles to manipulate the rate and structure of information spreading compared to traditional random, 0-Markov, circuits. For SWAP gates and brickwork circuits, we use graph cover time to demonstrate how $k$-Markov processes can be used to control operator transport. With SWAP gates and the set of Clifford gates that can change operator weight, we show how $k$-Markov sequences can be used to manipulate scrambling time and generate novel structures of spatial-temporal correlations across a qubit network. We show that $k$-Markov circuits constructed from PSWAP gates at fixed angle are equivalent to standard brickwork circuits with PSWAP angle drawn from non-uniform distributions generated by the $k$-Markov process. In those circuits, the time evolution of the average Hamming weight and the space-time correlation structure that remains after one-point quantities equilibrate again vary significantly from the 0-Markov case, depending on the transition probabilities of the process. More broadly, these results identify finite temporal memory as a control handle distinct from a circuit's degree of quantum chaos. For gate sets that fall short of volume-law scrambling, the equilibrium reached by one-point diagnostics is fixed by the gates alone, while the choice of $k$-Markov rule controls the approach to equilibrium and the structure of persistent, late-time correlations.
\end{abstract}

\maketitle


Understanding and controlling the dynamics of quantum correlations is a rich and active area of study. In particular, in large, many-body systems it is interesting to understand the classes of dynamics under which initially localized properties of a state remain localized, in contrast to those where local information becomes delocalized in correlations among many degrees of freedom. A large body of literature studies these questions in the context of quantum circuits, often with the evolution at each layer chosen at random from some specified set of operations that may include unitary gates \cite{Nahum:2018,Khemani:2018}, local measurements \cite{Li:2018,Skinner:2019MIPT,Chan:2019}, and dissipation \cite{Knap:2018,Schuster:2023,Weinstein:2023}. 

In this paper, we show how temporal correlations in a stochastic prescription for the unitary gates applied at each circuit layer can affect information dynamics. For circuits over a network of $N$ qubits, consider a fixed set of unitaries that may act on the network, each constructed from a tensor product of $p$-local unitaries, $p\leq N$. Labeling each $N$-qubit unitary by a letter ($A$,$B$, \dots) defines an alphabet from which a variety of stochastic sequences can be built, where the probability of a letter depends on $k$ previous letters (the context). For $k>1$, the sequences are $k$-Markov, often also called non-Markovian with context $k$. In contrast, most studies of random circuits to date assume $k=0$, or introduce temporal structure through measurement and feedback \cite{Raussendorf:2001,Zhang:2006,Buhrman:2024}. Here we show that $k$-Markov sequences provide operationally inexpensive dials to (statistically) control information dynamics without requiring mid-circuit disturbance of the state or real-time adjustment of the algorithm. In particular, the choice of sequence allows control of both the equilibration time of one-point quantities, and the structure of two-point correlations that persist to late times.

For a fixed set of $p$ available unitary gates, generating an example of any particular $k$-Markov rule is cheap: it requires storing a context of the previous $k$ letters and a lookup table of $(p-1)\,p^k$ conditional probabilities. A $k$-Markov rule can always be recast as $k=1$ by enlarging the alphabet from $p$ to $p^k$, where the $k\log_2 p$ bits that stored the context simply become the $\log_2(p^k)$ bits that label the current letter of the enlarged alphabet. In either case, $(p-1)\,p^k$ conditional probabilities are needed, the same $p$ physical gates must be available for each layer, and the process is not memoryless. Characterizing the ensemble generated by any one rule requires a volume of data that grows with the number of contexts, $p^k$, and is likewise invariant under recasting the process as 1-Markov. However, in the specific setting used here, where the alphabet corresponds to unitary gates, tradeoffs between Markov order and gate complexity are sometimes possible since the algebraic (multiplicative) properties of the gates may allow the sequences to be recast. As the examples below will illustrate, this means that the same statistical structure of correlations across a qubit network can sometimes, but not always, be generated by different combinations of gate set and process.

{\it Setting up the circuits - } We consider an alphabet of two letters ({\bf A}, {\bf B}), each defined by a particular choice of $N$-qubit unitary dynamics built from 2-local unitaries applied to nearest-neighbor qubits in a 1-dimensional, closed chain. For simplicity, we assume $N$ is even. From {\bf A} and {\bf B} we construct circuits using gate sequences determined by a $k$-Markov process. For example, $1$-Markov sequences are defined by two transition probabilities. These specify the probability that the next letter in the sequence is {\bf A}, conditioned on the previous letter being either {\bf A}, $p({\bf A}|{\bf A})$, or {\bf B}, $p({\bf A}|{\bf B})$. Similarly, a $2$-Markov rule is specified by four probabilities, $p({\bf A}|{\bf AA})$, $p({\bf A}|{\bf AB})$, $p({\bf A}|{\bf BA})$, $p({\bf A}|{\bf BB})$. (Of course, it is possible to choose probabilities at order $k$ such that the rule is actually lower order Markov. We exclude those cases from results shown at order $k$.) 

{\it Characterizing information dynamics - } To characterize the dynamics produced by the $k$-Markov gate sequences, we examine properties of $\mathcal{O}(\ell)=U^{\dagger}_{\ell}\mathcal{O}(0)U_{\ell}$, where $U_{\ell}$ is the cumulative unitary evolution of the circuit through layer $\ell$ and we take $\mathcal{O}(0)$ to be an operator with a non-identity component on only a single site. We will use the Pauli words, $\{\mathcal{P}_{\alpha}\}$, as the basis for operators and write $\mathcal{O}(\ell)=U^{\dagger}_{\ell}\mathcal{O}(0)U_{\ell}=\sum_{\alpha}c_{\alpha}(\ell)\mathcal{P}_{\alpha}$. The (Hamming) weight $w_\alpha$ of each word $\mathcal{P}_\alpha$ is its number
of non-identity components, and we quantify operator spreading through the
average weight $w(\ell)=\sum_\alpha |c_\alpha(\ell)|^2\, w_\alpha$, where the
coefficients satisfy $\sum_\alpha |c_\alpha(\ell)|^2=1$. For Clifford circuits
the evolved operator is a single Pauli word, so $w(\ell)$ is simply that word's
weight. While in general there are $4^N$ words, different choices of the allowed unitaries in the circuit and the initial operator can restrict the number of reachable Pauli strings, $R_{\rm reachable}(\mathcal{O}(0),\{U\})<4^{N}$, that can ever have non-zero weight in the evolved operator.

{\it SWAP gates and cover time - }We first demonstrate the relationship between $k$-Markov sequences and operator spreading in a simple case of only SWAP gates. With this dynamics, an initial single-site operator (we use $\sigma_z$) will simply move across the qubit network with time. The difference between dynamics can be quantified with cover time, $t_c$, which is the smallest number of circuit layers before every site has hosted $\sigma_z$ at least once. Consider the two-letter alphabet where the $N$-qubit unitaries are ${\bf A}_{\rm SWAP}$ and ${\bf B}_{\rm SWAP}$, corresponding to pairing each qubit to its left or right neighbor with a SWAP gate. Then, in the unbiased 0-Markov model, each layer of the circuit is chosen independently and with equal probability. The dynamics is an unbiased random walk on a one-dimensional cycle graph and $\langle t_c\rangle=\frac{N(N-1)}{2}$ \cite{lovasz:1993}. For simplicity, fix $P(A)=P(B)$. Then in each $k$-Markov model all probabilities will be symmetric under exchange of ${\bf A}$ and ${\bf B}$, but the mean and distribution of cover times will in general be quite different from those of the unbiased random walk. 

The shortest cover time in the SWAP gate case would be given by the (deterministic) sequence ${\bf A}_{\rm SWAP}{\bf B}_{\rm SWAP}{\bf A}_{\rm SWAP}{\bf B}_{\rm SWAP}...$ of length $N$. On the other hand, a sequence of only ${\bf A}_{\rm SWAP}$ or only ${\bf B}_{\rm SWAP}$ would localize the operator forever to just two sites. So, $k$-Markov processes will generate slower cover times when they bias the walk away from this optimal sequence. For $1$-Markov processes, the cover time will be correlated with the probability of optimal sequences (${\bf A}_{\rm SWAP}{\bf B}_{\rm SWAP}$ or ${\bf B}_{\rm SWAP}{\bf A}_{\rm SWAP}$), which is governed entirely by the probability to switch letters, $P_{\rm switch}=P(A|B)=P(B|A)$. Switching probabilities above (below) 0.5 will generate mean cover times that are shorter (longer) than that of the random walk. As $k$ is increased, the longer context allows choices that can help enforce the optimal trajectory.

The SWAP alphabet realizes an example where the sequence allows an algebraic collapse. Because ${\bf A}_{\rm SWAP}^2={\bf B}_{\rm SWAP}^2={\bf I}$, any
same-letter run reduces modulo two, so the cumulative evolution depends only on the parity-reduced, alternating word. The operator transport is therefore a persistent random walk \cite{Gilbert:2009}, where each step's direction is correlated with that of the previous step, so that the relevant state is the operator's position together with its current direction of motion. This state space is far smaller than the $2^k$ contexts of a generic order-$k$ rule. The memory nonetheless remains essential, controlling the ballistic-to-diffusive crossover in the cover time. The dynamics reduces to the memoryless $0$-Markov walk only at the isolated point $P_{\rm switch}=0.5$.

\begin{figure}[hbt]
\includegraphics[width=0.4\textwidth]{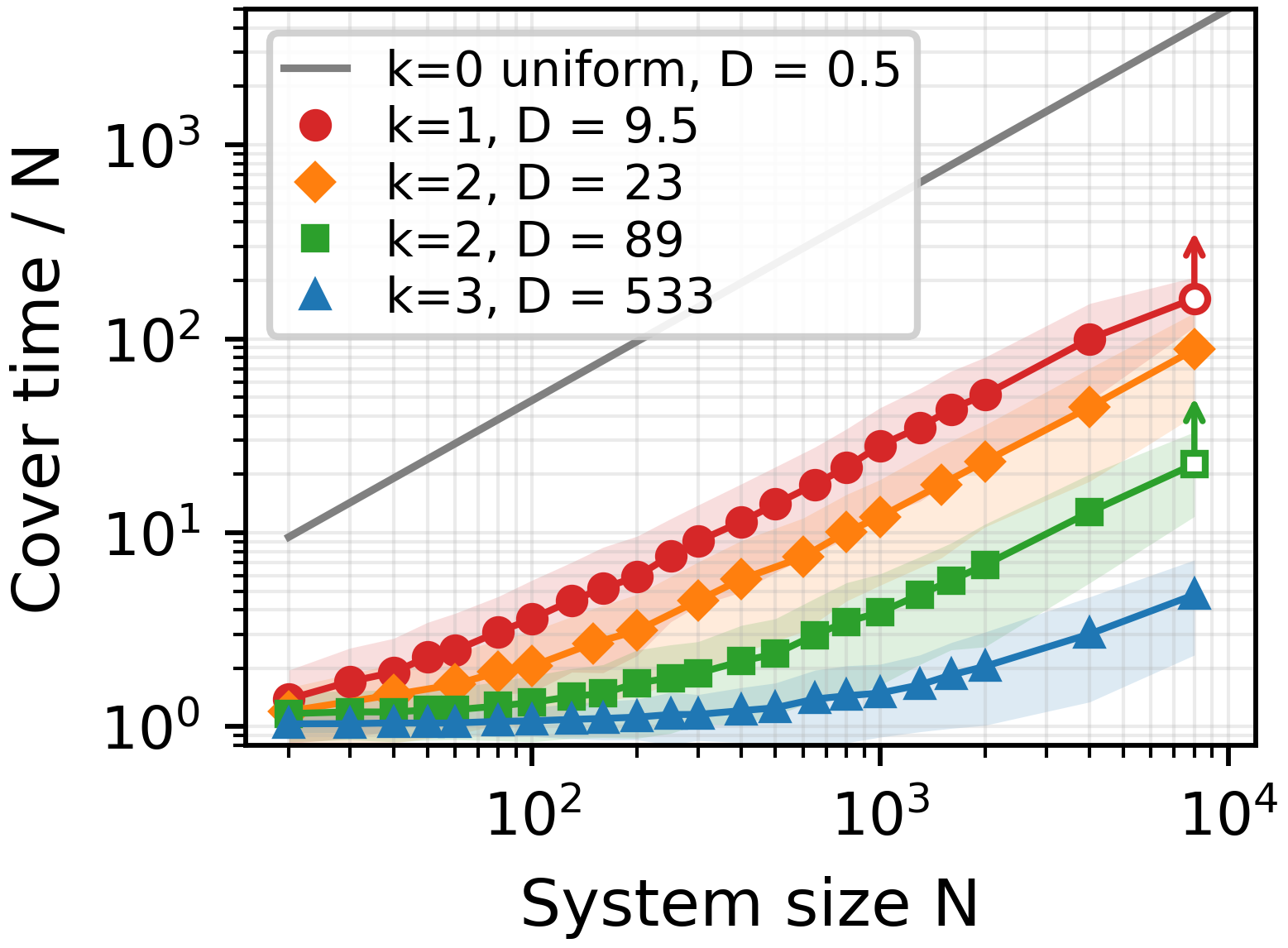}
\includegraphics[width=0.4\textwidth]{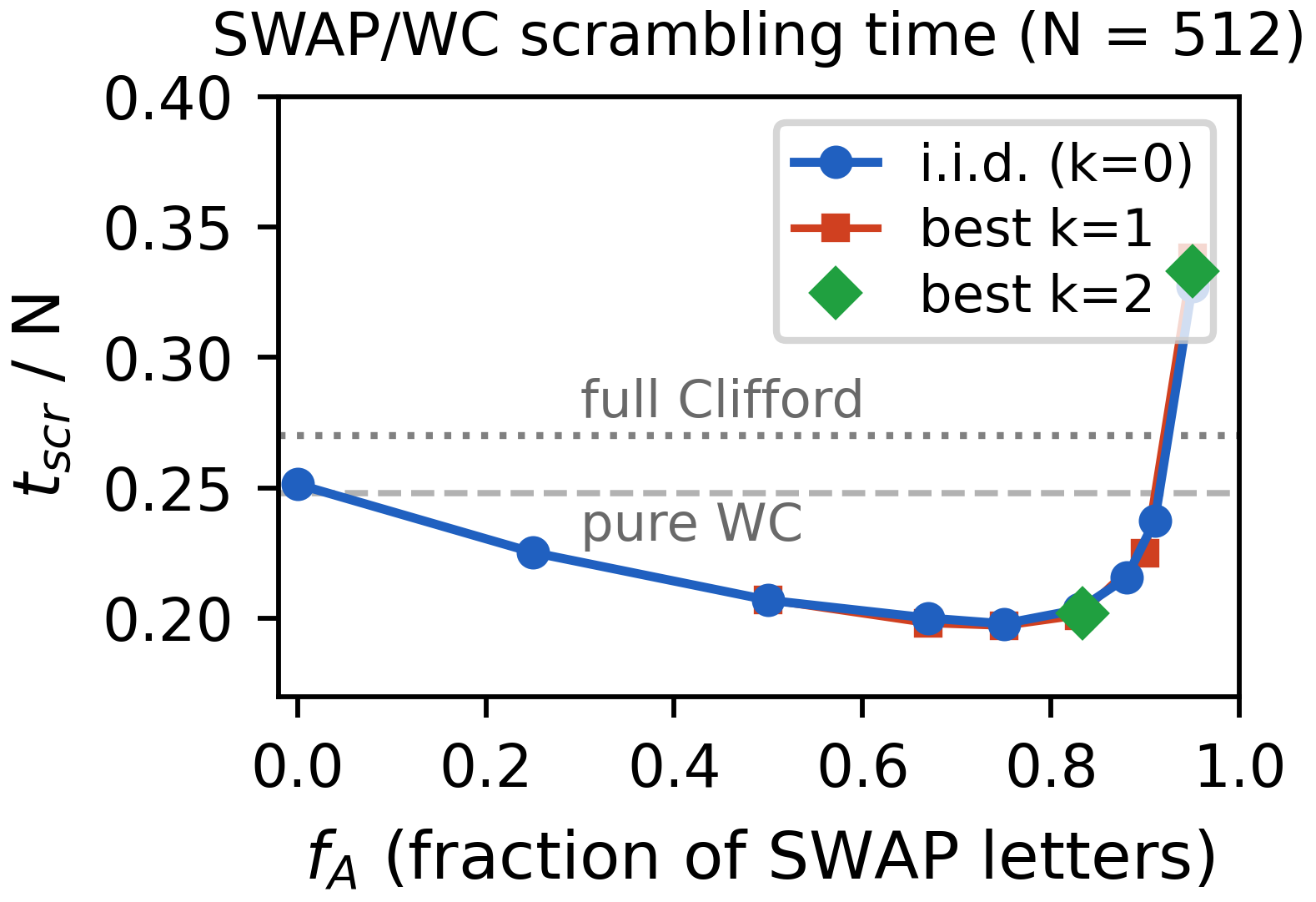}
\caption{\label{fig:CoverTimeN} {\bf Top:} Mean cover times over system size, $\langle t_c\rangle/N$ for $k$-Markov SWAP circuits as a function of $N$, for several different rules. The effective diffusion constant, $D$, is a function of the transition probabilities and controls the shift from nearly ballistic operator motion for smaller systems to diffusive transport in larger systems. Open symbols are used for cases where less than 90\% of runs reached full coverage before the limit $200N$ steps. For the $k=1$ example, only 54\% finished, while the $k=2$ example had 87\% finish. For those cases the position of the point is a lower bound on the mean cover time, biased low. Shaded regions show the $1\sigma$ variation over 500 trials. {\bf Bottom:} Scrambling times in Clifford weight-changing circuits, as a function of the fraction of SWAP single-layer (${\bf A}_{\rm SWAP}$), for $k=0$ rules and the shortest times found for $k=1$ and $k=2$, all for $N=512$.}
\end{figure}

The upper panel of Figure \ref{fig:CoverTimeN} illustrates this point, showing the mean cover time (and $\pm 1\sigma$ spread over 500 trials) of several $k$-Markov rules at varying network size, $N$. All rules share a common $P({\rm switch)}=0.95$, which significantly reduces cover time compared to the 0-Markov circuit. The dynamics of each process can be concisely described by its diffusion coefficient, $D$, which can be analytically computed from the transition probabilities in some cases \cite{Gilbert:2009}, or numerically computed by fitting the large-$N$ limit of each sequence ($t_C=N^2/4D$). For system size approximately $N<D$, the processes can maintain nearly ballistic motion of the operator, while for larger systems the errors from the optimal gate sequence build up sufficiently that the cover times again scale as $N^2$. Of course, different choices of the transition probabilities can lead to arbitrarily long cover times.

\begin{figure}[h]
  \centering
  \includegraphics[width=0.5\textwidth]{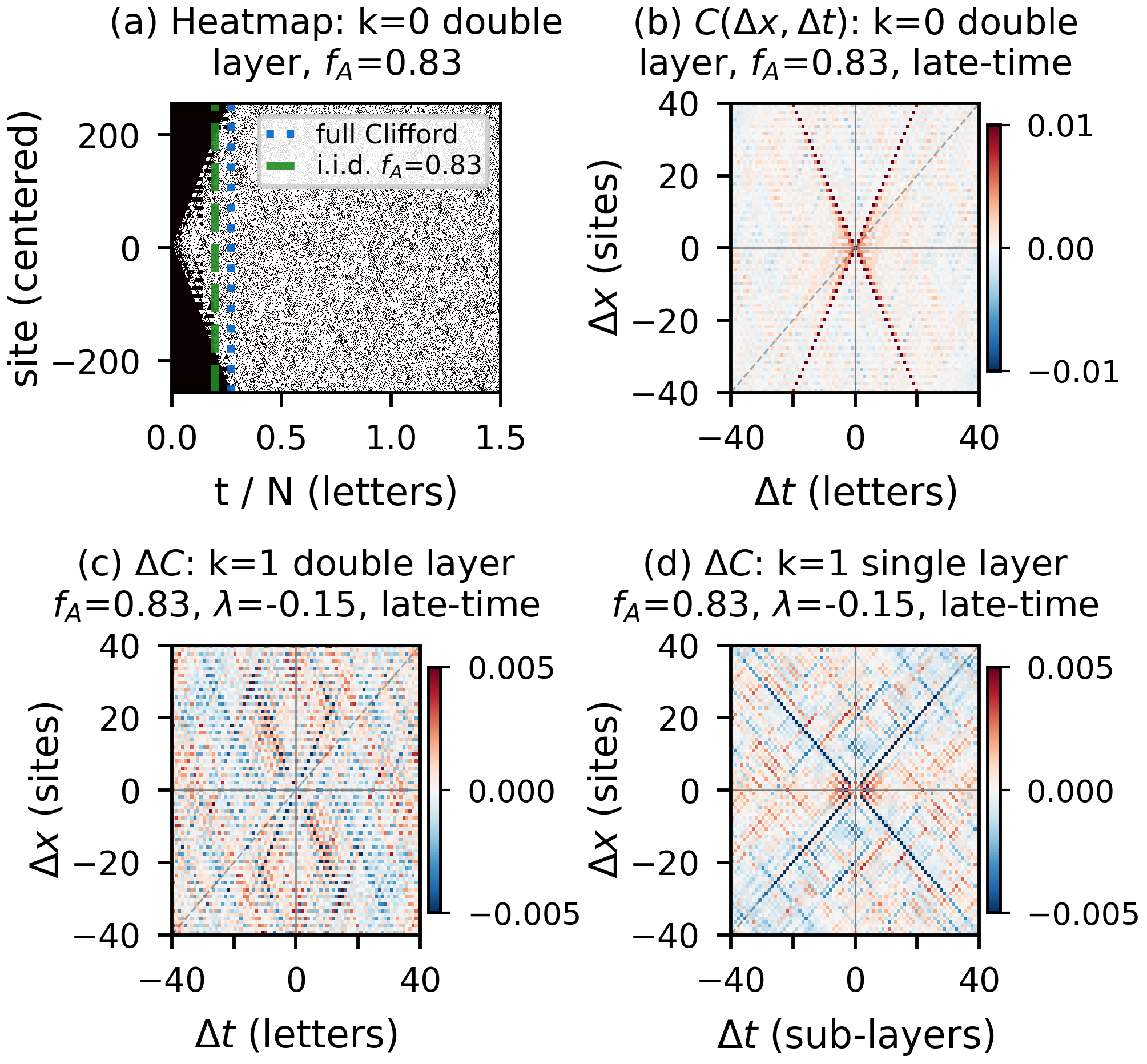}
  \caption{Panel (a) operator structure for one realization of a $k=0$ rule. The signatures of the SWAP gate transport are most visible before $t_{\rm scr}$, but partially retained even at very late times. (b) The late-time ($t\gg t_{\rm scr}$) correlation structure $C(\Delta x,\Delta t)$ averages over sites and 5 realizations. (c) and (d) The difference between double-layer and single-layer $k=1$ late-time correlations and those of $k=0$ at matched $f_A=0.83$. \label{fig:STcorrelations}}
\end{figure}

{\it Operator growth with Clifford gates - }
While in the previous example the operator weight was always 1, most unitary processes can change the weight, spreading the operator over multiple sites so that $w(\ell)$ grows. To consider what kind of dials the $k$-Markov processes can provide in that case, we next consider alphabets built of the SWAP gate (weight preserving) and the set of Clifford gates that can change operator weight. Under Clifford evolution $\sigma_z$ remains a single, signed Pauli string, and its support and weight are unchanged by overall sign flips. We therefore
only need the 720 Clifford operations that act distinctly on Pauli strings, rather than the full 11520. Of this smaller set, 648 cause transitions between at least one single-site operator and a two-site operator (increasing or decreasing the weight). A standard brickwork circuit with gates drawn at random from this set will increase operator weight faster thanks to the removal of the weight-conserving gates. However, the inclusion of SWAP gate layers can further accelerate growth by increasing the probability that the leading edge of non-trivial operator support moves outward. Consider fixed brickwork geometry, again an alphabet with two letters, ${\bf A}_{\rm SWAP}$ and ${\bf B}_{\rm WC}$. Now ${\bf A}_{\rm SWAP}$ applies SWAP gates to each qubit pair, while ${\bf B}_{\rm WC}$ chooses a gate uniformly at random from the weight-changing (WC) set. In contrast to consecutive ${\bf A}_{\rm SWAP}$ layers, which
cancel in pairs since ${\bf A}_{\rm SWAP}^2={\bf I}$, consecutive
${\bf B}_{\rm WC}$ layers are independent random draws whose products do not
simplify. A same-letter run of ${\bf B}_{\rm WC}$ approaches a uniformly random
Clifford within a couple of layers rather than reducing to a single effective
gate, so the weight-changing alphabet admits no algebraic collapse of the kind
available for SWAP or PSWAP, and the full $k$-context remains meaningful. Different dynamics can be achieved by choosing ${\bf A}_{\rm SWAP}$ and ${\bf B}_{\rm WC}$ to act on a consecutive pair of layers of brickwork (even and odd, called double-layer in Figure \ref{fig:STcorrelations}), or to break parity and act on a single layer (even or odd). 

To compare operator growth between models, we use the scrambling time, $t_{\rm scr}$, defined as the first layer for which the operator weight satisfies $w(\ell)\geq\frac{N}{2}$. For $k=0$ circuits, the scrambling time depends on the fraction of ${\bf A}_{\rm SWAP}$ gates, $f_A$. At $k=1,2$, we find that the scrambling time can only increase when ${\bf B}_{\rm WC}$, ${\bf A}_{\rm SWAP}$ each act on the even and odd bricklayers. When each letter is defined to act on only one layer (odd or even), only very small decreases in $\langle t_{\rm scr}\rangle$ can be achieved, where the optimal $f_A$ increases with $N$. The bottom panel of Figure \ref{fig:CoverTimeN} shows how $t_{\rm scr}$ scales with the fraction of single-layer ${\bf A}_{\rm SWAP}$ letters, $f_A$, for $k=0$ processes and $N=512$. It also shows that the shortest $t_{\rm scr}$ achieved with single-layer letters in scans over probabilities at fixed $f_A$ lie very close to the $k=0$ line. This near-coincidence reflects the absence of any algebraic collapse in the weight-changing alphabet.

While different $k$-Markov sequences may give the same mean scrambling time, they generically generate distinguishable spatial-temporal correlations that are strongest before $\langle t_{\rm scr}\rangle$, but that also persist at very late times. Figure \ref{fig:STcorrelations} shows several visualizations of the occupation at each site $i$: $n_i$=1 if the site hosts a non-identity operator, and is 0 otherwise. The figure shows a heatmap of occupation for 512 qubits (black is the identity operator, $n=0$) in panel (a). The remaining panels show different examples of the correlation function
\begin{equation}
\label{eq:correlation}
    C(\Delta x,\Delta t)=\frac{\langle n_i(x,t)n_i(x+\Delta x,t+\Delta t)\rangle-\langle n_i(x,t)\rangle^2}{\langle(n_i(x,t)-\langle n_i(x,t)\rangle)^2\rangle}
\end{equation}
where the average is over five realizations of each rule $\times$ 512 sites $\times\sim$1000 late-time steps (at $t > 2t_{\rm scr}$). The result is stable to $\sim0.1\%$ under changing or adding realizations. Panel (b) shows the late-time correlation function ($t\gg t_{\rm scr}$) for $k=0$, $f_A=0.83$ and double-layer letters. The lower two panels show the difference between the correlation function for $k=1$, $f_A=0.83$, $\lambda=1-\frac{P(A|B)}{1-f_A}$ and the $k=0$ correlation, for both single- and double-layer letters. Even at $k=0$, structure in $C$ from the SWAP gate transport is visible, but higher $k$ generates additional correlation.

 \begin{figure*}[t]
  \centering
\includegraphics[width=\textwidth]{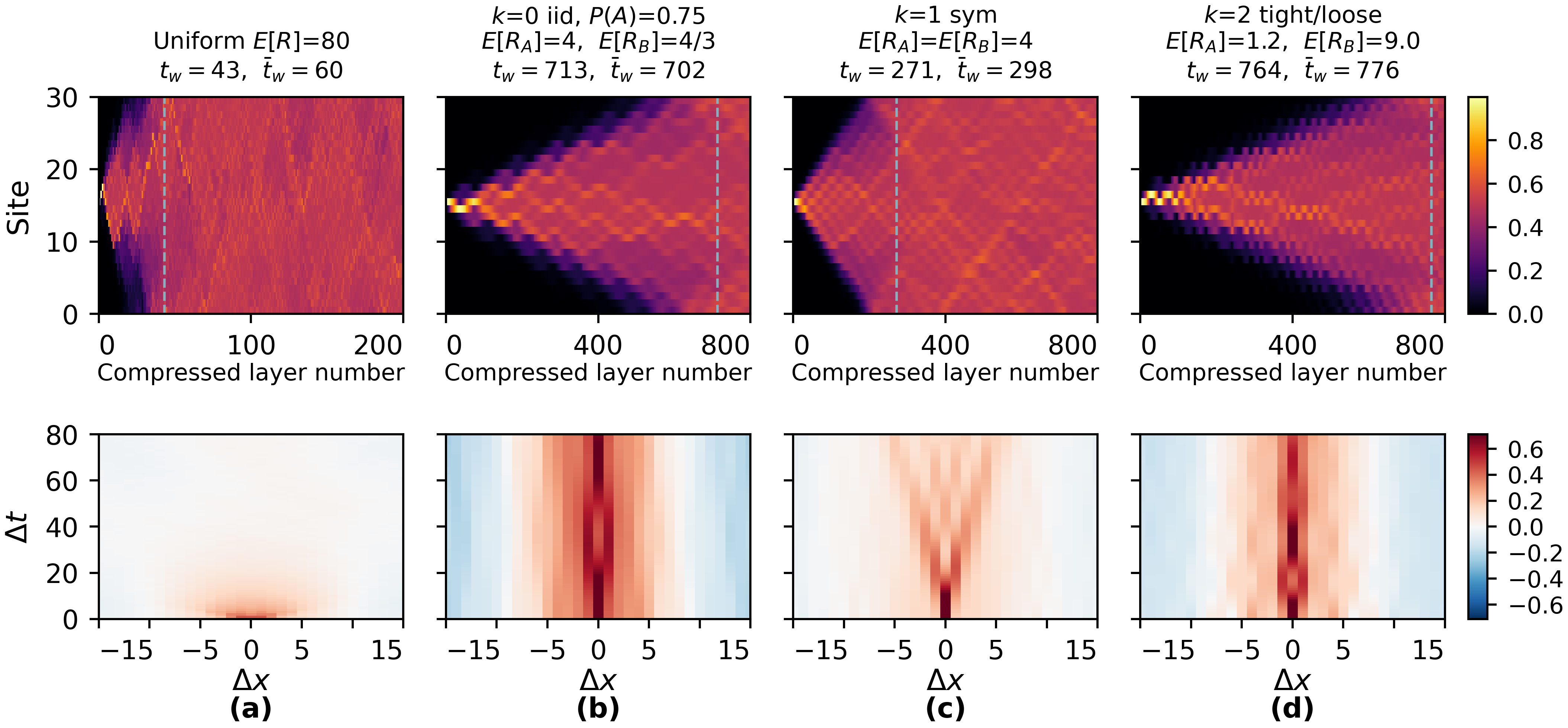}
  \caption{Heatmaps of the site occupation, $d_i$, (upper panels, one realization) and late-time spatial-temporal correlation structure of $d_i$ (lower panels, averaged over five realizations $\times$ 30 sites $\times \sim1000$ layers well after the network weight has equilibrated past the threshold) for a 30-qubit network, under four different rules for generating run-length distributions as follows {\bf (a)} Uniformly distributed, with average length $E[R_A]=E[R_B]=80$; {\bf (b)} Generated by a $k=0$ process for the $A$, $B$ alphabet, where $P(A)=0.75$ sets the geometric distributions $E[R_A]=4$, $E[R_B]=4/3$; {\bf (c)} Generated by a $k=1$ process where the geometric distributions for $R_A$, $R_B$ are chosen to be the same; {\bf (d)} Generated by a $k=2$ process with different expectation values for $A$ and $B$ run lengths. The result is stable to $\sim3\%$ under changing or adding realizations. \label{fig:eight_panel}}
 \end{figure*}

{\it PSWAP circuits - } Finally, we consider a class of circuits where operator weight and complexity can both grow, constructing the alphabet from two-qubit partial swaps (PSWAPs) with fixed angle $\theta$. We restrict PSWAP to act in the charge 1 sector only, since this reduces the number of Pauli strings reachable from $\sigma_z$ to $2N^2$ (for a 1-dimensional ring of qubits) and allows for a simpler analytic understanding of the effects. While Clifford circuits map Pauli strings to Pauli strings and thus significantly restrict the dynamics, PSWAP circuits map a Pauli string into a linear combination of Pauli strings.

As in the SWAP operator example, we take $\bf A_{\rm PSWAP}$ to be one of the two bricklayer couplings and $\bf B_{\rm PSWAP}$ to be the other. These PSWAP circuits can be directly connected to the large literature on random circuits that uses a brickwork pattern, drawing independent unitaries from a set for each layer \cite{Khemani:2018}. This is because a run of $R$ consecutive ${\bf A}_{\rm PSWAP}$ (or ${\bf B}_{\rm PSWAP}$) layers collapses into a single layer, since the two-qubit unitaries satisfy $\left(U_{i,i+1}(\theta)\right)^{R}=U_{i,i+1}(R\theta)$. The circuits now have standard brickwork geometry, but where the evolution at compressed layer $\ell$ is given by
\begin{equation}
\label{eq:collapsedU}
U(\ell)=\prod_j U(R_{A,\ell}\theta)_{2j-1,2j}\prod_j U(R_{B,\ell}\theta)_{2j,2j+1} \,,
\end{equation}
where $R_{A,\ell}$ ($R_{B,\ell}$) is the length of the $\ell$-th ${\bf A}$-run (${\bf B}$-run), and $R_A,R_B$ denote the corresponding random variables. These are brickwork circuits with each $R_{A,\ell},R_{B,\ell}$ drawn from distributions inherited from the $k$-Markov rule. 

Unbiased, $0$-Markov circuits give $P(R_A)=P(R_B)=2^{-R}$; a biased $0$-Markov rule ($P({\bf A})\neq\tfrac12$) still gives independent geometric runs, but the
$A$- and $B$-run means, $1/P({\bf B})$ and $1/P({\bf A})$, are both fixed by the single letter probability and so cannot be chosen separately. At $k=1$ the two switch probabilities decouple the run length distributions, giving $E[R_A]=1/P({\bf B}|{\bf A})$ and $E[R_B]=1/P({\bf A}|{\bf B})$. We show details of the calculations in Appendix~\ref{app:runlength}, and here quote just the key qualitative features. At $k=2$, the distributions for angle are no longer geometric, since the probability to enter either same-letter run need not be simply related to the probability of staying in the run \cite{balakrishnan2002runs}.

At both $k=1$ and $k=2$ the successive run lengths remain statistically independent, so the compressed circuit is a genuinely memoryless ($k=0$) brickwork circuit over a distribution of discrete angles $R\theta$, geometric at $k=1$ and non-geometric at $k=2$. This is an example where an order-versus-gate-complexity tradeoff is possible, replacing a single fixed-angle gate together with run-length memory by independent draws from a broader set of angles. At $k\geq 3$, the distributions for temporally separated same-letter run lengths start to be correlated, as long as typical run lengths are shorter than $k$. While recent work has considered information dynamics in 0-Markov brickwork circuits when unitaries drawn from a non-Haar distribution \cite{Zhiyang:2025}, these $k$-Markov circuits suggest a large class of related studies.

As we found with the scrambling time for the weight-changing Clifford circuits, the standard single-time diagnostics of operator spreading classify circuits by measures whose equilibrium values depend on the gate set alone, and not the $k$-Markov rule. For example, the out-of-time-order correlator, averaged over local Pauli probes, equals (up to a
constant) the mean occupation of a site. It is a first moment of the occupation field, and summed over sites is $\tfrac43\,w(\ell)$~\cite{Nahum:2018,Khemani:2018}. The operator-space
entanglement entropy across a spatial cut behaves likewise, growing with the weight and saturating to a small, rule-independent, sub-volume value set by the free-fermion structure of the
charge-1 PSWAP \cite{Jozsa:2008, Prosen:2007}. Because these are all first-moment (mean) quantities, they cannot register the temporal correlations of the drive. Those will appear in the connected two-point function of the occupation and can persist after the mean has equilibrated.

Figure~\ref{fig:eight_panel} shows several examples for 30 qubits, for which the equilibrium weight is empirically $w/N\approx 0.52$--$0.53$. The run length statistics quoted were computed using the equations given in Appendix~\ref{app:runlength}. When the physical angle applied in each unitary is small (we use $\theta=\pi/300$), short same-letter runs (lengths $R$) move weight of about ${\rm sin}^2 [R\theta]\approx (R\theta)^2$ to operators with support on the neighboring qubit, generating a smooth light cone. Equilibration in $w/N$ is fastest, and smooth, when run lengths cluster at the longest length such that $R\theta\lesssim1$. For these examples we report $t_w$, the first compressed step where $w/N \geq 0.50$, for individual realizations and the average over many trials. When longer run-lengths can occur, the smooth spreading is punctuated by bursts that are large enough that $R\theta$ is not small. As in the previous section, many different evolutions can give a comparable equilibration time to $w/N$, but the structure in the light cone and the correlation structure after equilibration will remain distinct. The quantity shown in the heatmaps is now the average occupation at site $i$, $d_i=\sum_{\alpha:\mathcal{P}_{\alpha,i}\neq I}|c_{\alpha}|^2$, where only operators that are not the identity at site $i$ are included. The correlation panels are made with $d_i$ in Eq.(\ref{eq:correlation}) (instead of $n_i$) and illustrate the persistent late-time distinctions between rules: the broad uniform distribution (a) leaves almost no late-time correlation, the short geometric $k=0$ runs (b) produce a strong equal-site ($\Delta x=0$) temporal correlation with anticorrelated flanks, the $k=1$ rule (c) instead shows correlation along the ballistic light-cone diagonals, and the asymmetric $k=2$ rule (d) adds a pronounced temporal banding. The occupation $d_i(\ell)$ is the operator weight at site $i$, and its sum over sites is the average Hamming weight, $\sum_i d_i(\ell)=w(\ell)$. The average over the three probe operators $\{X_i,Y_i,Z_i\}$ at site $i$ of the infinite-temperature OTOC, $C=\tfrac12\langle|[\mathcal{O}(\ell),V]|^2\rangle$, equals $\tfrac43\,d_i(\ell)$~\cite{Nahum:2018,Khemani:2018,Hosur:2016}, so the occupation heatmaps are OTOC light cones. The cover- and scrambling-time diagnostics from previous subsections were built from $w=\sum_i d_i$, and so are also OTOC-front quantities.

\begin{figure}[h]
  \centering
\includegraphics[width=0.5\textwidth]{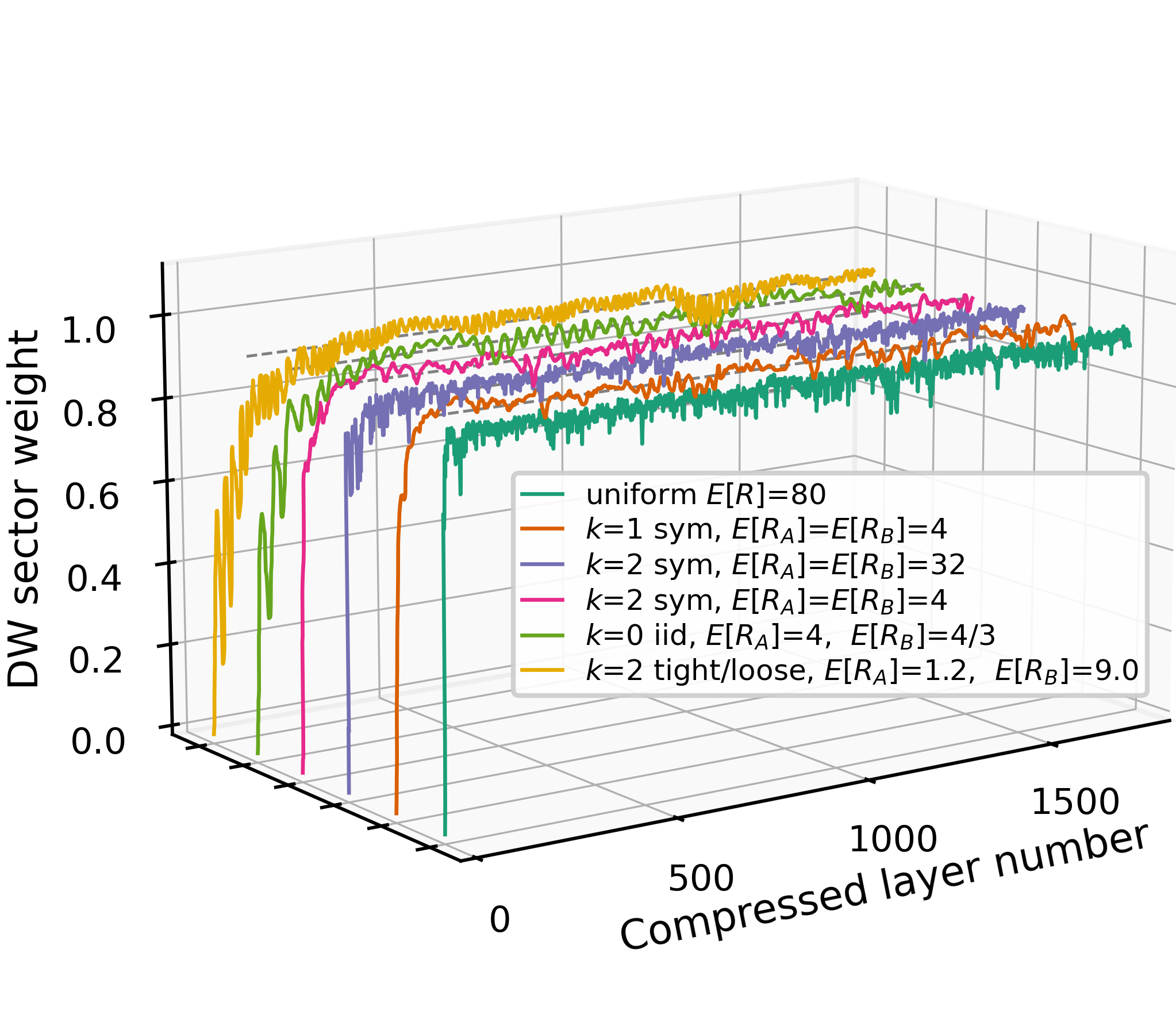}
  \caption{The time evolution of the weight in the domain-wall sector of operators, for a single realization of each model scenario shown in Figure \ref{fig:eight_panel} and two others. The sector weights satisfy $w_Z+w_{\rm DW}=1$ at all layers, so $w_{\rm DW}$ is the fraction of the total operator norm in the domain-wall sector. All models approach a common late-time value $w_{\rm DW}\approx0.90$, a property of the operator orbit (its dephasing equilibrium) rather than of the $k$-Markov rule. This value is shown as a dashed line under each curve, with curves offset along the depth axis for visibility). The models differ in the transient and in the frequency content of the oscillations about this shared asymptote. The $k=2$ symmetric case with $E[R_A]=E[R_B]=4$ shows that at small run length (and for the same seed), the difference between the geometric distributions at $k=1$ is small. The $k=2$ symmetric case with $E[R_A]=E[R_B]=32$ shows the different effective filter applied by distributions that contain more long run lengths. \label{fig:DWweight}}
\end{figure}

Some properties of the gate sequence statistics can be understood by dividing the accessible Pauli strings into two sectors: $2N$ operators with only $\sigma_z$ (and identity) operators, and $N(2N-2)$ 'domain wall' strings where a single pair of $\sigma_x$, $\sigma_y$ operators is separated by a string of $\sigma_z$ operators. The initial operator is entirely in the $Z$-sector, and each PSWAP gate moves some weight into the domain wall sector. The weight in the domain wall sector is $w_{\rm DW} =\sum_{\alpha:\,n_{XY}(\mathcal{P_{\alpha}})=2} |c_{\alpha}|^2$. The time-evolution of this weight is shown for several different models in Figure \ref{fig:DWweight}. The figure shows that while each rule approaches late-time mean value $w_{\rm DW}\approx0.90$ (indicated with dashed lines in the figure), different run-length statistics, generated by different $k$-Markov models, act as filters for the oscillation frequencies visible in the weight evolution.

These oscillations arise from coherent exchange between the two sectors. Each PSWAP layer rotates a small amount of weight between neighboring $Z$-strings and domain-wall pairs, and the pairs, once created, propagate ballistically and partially recombine, so weight flows back and forth between the sectors. In the eigenbasis of the brickwork evolution these motions occupy a band of frequencies of order the rotation rate $2\theta$, and the sector-weight oscillations appear at differences of these frequencies. Because a same-letter run of length $R$ applies $U(R\theta)$, the run-length distribution set by each $k$-Markov rule samples these coherent oscillations at different intervals and thereby damps them selectively, most strongly at high frequencies and for distributions weighted toward long runs. Different rules therefore leave distinct spectral signatures in the weight evolution even though the equilibrium sector populations they reach are identical.

This combination of a common asymptotic domain-wall weight with rule-dependent oscillations mirrors the results found for the weight-changing Clifford circuits of the previous section. There, sequences with comparable scrambling times still produced distinguishable spatial-temporal correlations that persisted well after equilibration. Here too, the choice of $k$-Markov rule leaves the equilibrium sector populations unchanged while imprinting lasting structure in the sector-weight oscillations and the
late-time correlations (Fig.~\ref{fig:eight_panel}, lower panels).

{\it Summary and Discussion- } We have demonstrated that $k$-Markov sequences of unitary gates can be used to generate families of stochastic circuits that display a variety of structures in operator evolution. This very broad class of models has two components: the definition of an alphabet, constructed from some choice of unitary dynamics on a qubit network, and the specification of transition probabilities defining a $k$-Markov process on the alphabet. 

Beyond the comparison with random, $k=0$, circuits, the stochastic sequences considered here can be contrasted with the deterministic sequences that have been explored for structured driving~\cite{Nandy:2017,Pilatowsky:2023,Morphic:2025}. While $k$-Markov sequences have high factor complexity (every length $n$-block can occur) with a tunably decaying autocorrelation, deterministic sequences allow low complexity with a persistent autocorrelation whose structure can be tuned. For example, with deterministic two-letter sequences the autocorrelation can be periodic (Floquet sequences~\cite{Bukov:2015,Bertini:2019}), quasi-periodic (Fibonacci words~\cite{Pilatowsky:2023}), or self-similar (Thue-Morse substitution words~\cite{Nandy:2017,Morphic:2025}). The long-range order of these deterministic drives has been demonstrated to generate emergent Hilbert-space ergodicity and randomness~\cite{Pilatowsky:2023,Morphic:2025} or control heating toward prethermal steady states~\cite{Nandy:2017}.

The temporal memory of the $k$-Markov sequences allows a type of control that is distinct from the quantum chaos axis. The three gate families used in this work all lie outside the volume-law quantum chaos family, for distinct reasons: SWAP transports operator weight without generating it, the weight-changing Clifford gates map Pauli strings to Pauli strings and so do not generate any operator entanglement, and the charge-1, PSWAP is a free-fermion (matchgate) gate so the operator-space entanglement stays $O(1)$ (sub-volume). For these examples, there is a gate-set-determined equilibrium value for one-point quantities (the covered configuration, the Hamming weight plateua, and the domain-wall sector weight), that is independent of the gate sequence. The $k$-Markov sequence controls the rate of approach (the cover time, scrambling time, and weight-equilibration time) and the structure of higher-order correlations that distinguish different $k$-Markov models. The correlations persist without decay for the longest time we've checked, out to about $10^6$ layers for the PSWAP circuits. We expect the control enabled by the $k$-Markov gates to weaken if the gate set used approaches that needed for rapid volume-law scrambling, where the self-averaging property of those circuits will tend to wash out the correlations. However, for gate sets where operator complexity remains low, we also expect that when applied to states, rather than operators, the $k$-Markov sequences may slow, route, or revive local quantum information dynamics relative to a memoryless gate sequence~\cite{Akhouri:2025pdz}.

{\it Acknowledgments - } We thank Zhen Bi, Tommy Chin, Stefan Eccles, Jeysen Flores-Velazquez, Tom Iadecola, Dezhe Jin and Xiantao Li for helpful discussions. The work of Pei-Jun Huang was supported by the Penn State NSF REU through PHY-2349159. This work was supported by the NSF through PHY-2310662, and with computational resources supported by the NSF through OAC-2201445.

\appendix

\section{Run-length distributions of $k$-Markov sequences}
\label{app:runlength}
For completeness we collect the run-length statistics used in the main text;
they are elementary consequences of the Markov property (for the broader theory
of runs and patterns, see \cite{balakrishnan2002runs}). A \emph{run} is a maximal
block of identical letters, so each of the $k$-Markov sequences considered here is composed entirely of alternating ${\bf A}$ and ${\bf B}$ runs.

To compute the statistics of run length, let $p_j$ be the probability that a run continues past its $j$-th symbol (i.e. that the $(j{+}1)$-th letter equals the $j$-th). Then the run length has distribution
\begin{equation}
P(R=r)=\Big(\prod_{j=1}^{r-1}p_j\Big)(1-p_r),\qquad r\ge1,
\end{equation}
with mean
\begin{equation}
\label{eq:ER}
E[R]=\sum_{r=1}^{\infty} r\,P(R=r)=\sum_{r=0}^{\infty}\prod_{j=1}^{r}p_j\,.
\end{equation}
To obtain the second equality we have used the tail-sum identity, $E[R]=\sum_{r\ge1}P(R\ge r)$, and
$P(R\ge r)=\prod_{j=1}^{r-1}p_j$ (with $P(R\ge1)=1$).
For a $k$-Markov rule $p_j$ takes at most $k$ distinct values since the decision depends on the last
$k$ letters, which are all identical once the run has extended $k-1$ steps. Then
$p_j$ equals a fixed ``bulk'' value for $j\ge k$, and a closed form of Eq.~\eqref{eq:ER} can be obtained once the probabilities of the sequence have been specified.

From the general case, we can see the properties of low-$k$ run distributions used in the text. For $k=0$ and $k=1$, the continuation probability is constant, $p_j\equiv p$.
For an ${\bf A}$-run $p=P({\bf A}|{\bf A})=1-P({\bf B}|{\bf A})$ (with $P({\bf A}|{\bf A})=P({\bf A})$ at $k=0$), so $P(R=r)=p^{r-1}(1-p)$ is geometric and runs are independent. Eq.~\eqref{eq:ER} gives $E[R_A]=1/(1-p)=1/P({\bf B}|{\bf A})$ and $E[R_B]=1/P({\bf A}|{\bf B})$. The unbiased case $P({\bf A})=\tfrac12$ gives $P(R)=2^{-R}$,
$E[R]=2$.

At $k=2$, every ${\bf A}$-run is entered from the fixed context $({\bf B},{\bf A})$. This means that consecutive runs remain
independent, but the first continuation probability differs from the bulk probability:
$p_1=P({\bf A}|{\bf BA})$ and $p_{j\ge2}=P({\bf A}|{\bf AA})$. Then
\begin{align}
P(R_A{=}1)&=1-P({\bf A}|{\bf BA})\,,\nonumber\\
P(R_A{=}r{\ge}2)&=P({\bf A}|{\bf BA})\,[P({\bf A}|{\bf AA})]^{r-2}[1-P({\bf A}|{\bf AA})],
\end{align}
and Eq.~\eqref{eq:ER} gives
\begin{equation}
E[R_A]=1+\frac{P({\bf A}|{\bf BA})}{1-P({\bf A}|{\bf AA})},
\end{equation}
with the analogue for $R_B$ under ${\bf A}\leftrightarrow{\bf B}$.

At $k\ge3$, the independence of runs starts to be lost. Now the entry context can reach across a previous switch between run type, since for example an
${\bf A}$-run at $k=3$ can be entered from $(\dots,{\bf A},{\bf B},{\bf A})$ or
$(\dots,{\bf B},{\bf B},{\bf A})$ according to whether the preceding ${\bf B}$-run
had length $1$ or ${\ge}2$. The $p_j$ then depend on the previous run(s) and
consecutive run lengths are generically correlated if typical runs are shorter than
$k$.

\providecommand{\noopsort}[1]{}\providecommand{\singleletter}[1]{#1}%

\end{document}